\documentclass[journal=ancac3,manuscript=article]{achemso}
\usepackage{chemformula} 
\usepackage[T1]{fontenc} 
\usepackage{graphicx}
\usepackage{fancyhdr}
\usepackage{makeidx}
\usepackage{mfirstuc}
\usepackage{physics}
\usepackage{upquote}
\usepackage[normalem]{ulem}
\usepackage{xcolor}
\usepackage{comment}
\usepackage{enumerate}
\usepackage{pdfpages}
\graphicspath{{./Figures_v9/}}
\usepackage{amssymb,amsmath,amsfonts}
\usepackage[figurename=Figure]{caption}
\usepackage[labelfont=bf]{caption}
\usepackage{xr}
\setkeys{acs}{maxauthors = 0}
\setkeys{acs}{keywords = true}

\newcommand{\bs}{\boldsymbol}

\author{Medha Dandu}
\affiliation[IIScECE]
{Department of Electrical Communication Engineering, Indian Institute of Science, Bangalore 560012, India}
\author{Garima Gupta}
\affiliation[IIScECE]
{Department of Electrical Communication Engineering, Indian Institute of Science, Bangalore 560012, India}
\author{Pushkar Dasika}
\affiliation[IIScECE]
{Department of Electrical Communication Engineering, Indian Institute of Science, Bangalore 560012, India}
\author{Kenji Watanabe}
\affiliation[RCFMNIMS]
{Research Center for Functional Materials, National Institute for Materials Science, Namiki 1-1, Tsukuba, Ibaraki 305-0044, Japan.}
\author{Takashi Taniguchi}
\affiliation[ICMNNIMS]
{International Center for Materials Nanoarchitectonics, National Institute for Materials Science, Namiki 1-1, Tsukuba, Ibaraki 305-0044, Japan.}
\author{Kausik Majumdar}
\email{kausikm@iisc.ac.in}
\affiliation[IIScECE]
{Department of Electrical Communication Engineering, Indian Institute of Science, Bangalore 560012, India}

\begin{keywords}
{twisted bilayer, moir\'e, excitons, trions, valley coherence, raman scattering }
\end{keywords}
\title{Electrically Tunable Localized \textit{versus} Delocalized Intralayer Moir\'e Excitons and Trions in a Twisted MoS$_2$ Bilayer}

\begin{document}
\newpage
{\abstract Moir\'e superlattice-induced sub-bands in twisted van der Waals homo- and hetero-structures govern their optical and electrical properties, rendering additional degrees of freedom such as twist angle. Here, we demonstrate the moir\'e superlattice effects on the intralayer excitons and trions in a twisted bilayer of MoS$_2$ of H-type stacking at marginal twist angles. We identify the emission from localized and multiple delocalized sub-bands of intralayer moir\'e excitons and show their electrical modulation by the corresponding trion formation. The electrical control of the oscillator strength of the moir\'e excitons also results in a strong tunability of resonant Raman scattering. We find that the gate-induced doping significantly modulates the electronic moir\'e potential, however leaves the excitonic moir\'e confinement unaltered. This effect, coupled with variable moir\'e trap filling by tuning the optical excitation density, allows us to delineate the different phases of localized and delocalized moir\'e trions. We demonstrate that the moir\'e excitons exhibit strong valley coherence that changes in a striking non-monotonic W-shape with gating due to motional narrowing. These observations from the simultaneous electrostatic control of quasiparticle-dependent moir\'e potential will lead to exciting effects of tunable many-body phenomena in moir\'e superlattices.}
\section*{Keywords}
twisted bilayer, moir\'e, excitons, trions, valley coherence, raman scattering

\newpage
The creation of artificial superlattices by stacking layered materials with precise control of twist angle has propelled a new research field of twistronics\cite{hennighausen2021}, offering numerous possibilities to engineer and simulate various physical phenomena\cite{carr2020,tang2020}. For example, such superlattices host diverse quantum phases of correlated electron behaviour such as high-transition-temperature superconductivity\cite{chen2019,balents2020} and Wigner crystallization\cite{regan2020} with the evolution of extremely flat-bands\cite{vitale2021}. In aligned heterobilayers of transition metal dichalcogenides (TMDs), emergence of moir\'e patterns\cite{he2021} with large period and deep lattice potential\cite{guo2020} results in spatial modulation of the optical selection rules\cite{yu2017}, sub-bands\cite{bremtun2020,ruiz2020} and trapping\cite{tran2019,seyler2019} of interlayer (IX) excitons. Twisted TMD homobilayers are also emerging as an interesting platform for the study of moir\'e superlattice (mSL) effects. Robust control of twist angle through `tear and stack' technique\cite{kim2016} eases the preparation of reliable homobilayer samples providing access to larger moir\'e length scales. Recent experiments reveal atomic reconstruction\cite{weston2020}, ultra-flat-bands\cite{li2021}, electrically tunable hybrid excitons\cite{sung2020}and valley dynamics\cite{scuri2020,andersen2021} in twisted TMD homobilayers.
\\
\\
In this work, we focus on the mSL effects on the intralayer excitons and their charged complexes in a twisted bilayer (tBL) of MoS$_2$. Fig. 1(a) illustrates the mSL of a tBL-MoS$_2$, formed by H-type stacking of two monolayers with a twist angle, $\theta$. The periodicity of the mSL is given by $a_{M} = \dfrac{a}{2sin(\theta/2)}$ where $a$ is the lattice constant of MoS$_2$. The spatial variation of the atomic registries in an mSL induces a corresponding periodic modulation in the interlayer spacing and coupling which in turn results in a periodic mSL potential for electrons ($V^{e}_{M}$) and holes ($V^{h}_{M}$) as illustrated in Fig. 1(b) for the conduction (CB) and valence band (VB) minima at the $\bs{K}$ point (see Note-I in SI).  In an H-type stacked tBL, $H^{M}_{X}$ and $H^{M}_{M}$ regions [denoted by A and B in Fig. 1(a)] feature global and local potential minima respectively\cite{shabani2021}. Bragg reflection of the carriers from the periodic moir\'e potential leads to a folding of the original electronic bands into a series of closely spaced sub-bands in the mini Brilluoin zone (mBZ).
\\
\\
Similarly, the intralayer exciton bands also fold into multiple sub-bands in mBZ, whose energy spacing is governed by the Fourier interlayer hopping terms at the $\bs{K}$ valleys. The excitons with near-zero center of mass (COM) momentum in multiple sub-bands become optically active in a twisted bilayer resulting in redistribution of the oscillator strength of the exciton of the corresponding monolayer. The moir\'e potential ($V^{ex}_{M}$) for COM motion of the intralayer excitons at the $\bs{K}$ valley is illustrated in Fig. 1(c). Despite the shallower $V^{ex}_{M}$\cite{yu2017}, recent theoretical predictions reveal that the intralayer excitons in TMD tBLs also feature multiple emission peaks\cite{wu2017,bremhyb2020} for twist angle $\leq 2^{0}$. Splitting in intralayer exciton peaks is also experimentally observed in closely aligned heterobilayer stacks of WS$_2$/WSe$_2$\cite{jin2019} and MoSe$_2$/MoS$_2$\cite{zhang2018}. Even at very small $\theta$, only the lowest energy sub-band of intralayer exciton exhibits flat-band nature due to shallower $V^{ex}_{M}$ while the interlayer excitons exhibit multiple flat sub-bands due to deeper potential\cite{bremtun2020}. Such flat-band nature localizes excitons like that in a quantum dot due to vanishing group velocity that inhibits hopping between the adjacent moir\'e wells. Top panel of Fig. 1(c) illustrates the moir\'e localized ($X_{L}$) and delocalized ($X_{D}$) intralayer exciton sub-bands. Fig. 1(c) highlights the cases of low (middle panel) and high intensity (bottom panel) optical excitation in a tBL and the corresponding distribution of the intralayer excitons. At a high intensity excitation, the subsequent filling of moir\'e traps is expected to lead to a saturation of the emission from $X_{L}$. The density of moir\'e traps ($N_{m}$) is about $\sim 0.77\:(3.07) \times10^{11}\:cm^{-2}$ for $\theta = 0.5^{0} (1^{0})$. In our measurement setup, the optical excitation density is $\sim 10^{3}$-fold higher in pulsed mode ($\sim 10^{12}\:cm^{-2}$) relative to that of continuous wave (CW) laser and this helps in reaching exciton density higher than $N_{m}$.
\\
\\
In our study, we provide the experimental evidence of the co-existence of such localized and delocalized intralayer moir\'e excitons using  tBL-MoS$_2$ samples in the dual-gated device architecture shown in Fig. 1(d) [see Methods for fabrication details]. We first apply the gate voltages and then optically excite the sample to collect the photoluminescence (PL). This enables the injected carrier concentration to reach its equilibrium distribution across the sample before optical excitation. We simultaneously apply the top ($V_{tg}$) and bottom ($V_{bg}$) gate voltages in the ratio of top and bottom hBN thickness (all the results are presented here as a function of $V_{bg}$). All the measurements are performed at 4.5 K unless otherwise mentioned. We measured a total of six different tBL-MoS$_2$ devices with H-type stacking - D1 (near-$0^{0}$), D3 (near-$1^{0}$), D4 (near-$3^{0}$), D5 (near-$0^{0}$), D6 (near-$1^{0}$) and D7 (near-$1^{0}$), and an nBL-MoS$_2$ device - D2. All the results from tBL-MoS$_2$ discussed in the main text correspond to the device D1 [see Fig. 1(e)].
\section*{Results and Discussion}
\textbf{Emission from intralayer moir\'e exciton sub-bands:} First, we probe the steady-state PL emission of tBL-MoS$_2$ with 532 nm CW laser, under floating gates. The PL spectrum from tBL region of D1 clearly exhibits multiple peaks around the A$_{1s}$ exciton regime of MoS$_2$ denoted by $X_{I}$, $X_{II}$, $X_{III}$, and $X_{IV}$, as shown in the top panel of Fig. 2(a). This is in contrast to the PL spectra from natural bilayer (nBL) of device D2 (middle panel) and monolayer (1L) region of D1 (bottom panel). The corresponding optical image of D2 is shown in Fig. S2(a). The lower energy peaks, $X_{I}$ and $X_{II}$, exhibit a red-shift and the higher energy peaks, $X_{III}$ and $X_{IV}$, evolve with a blue-shift relative to the A$_{1s}$ exciton of 1L-MoS$_2$. Interestingly, $X_{II}$ peak exhibits much higher intensity than the lowest energy peak, $X_{I}$.
\\
\\
Further, we observe that by slightly increasing $\theta$, the red-shift of $X_{I}$ decreases and the higher energy peaks, $X_{III}$ and $X_{IV}$ disappear (see Fig. S3 for PL comparison of D1, D3 and D4). These results are in good agreement with the theoretical calculations of optical conductivity spectra of tBLs\cite{bremhyb2020,wu2017}, and ascertain the formation of exciton sub-bands in mSL of tBLs, enabled by small $\theta$. The near-$0^{0}$ H-type stacking of tBL in D1 can be confirmed from the similar peak positions of both $A_{1g}$ and $E_{2g}^{1}$ Raman modes\cite{grzeszczyk2021} in tBL and nBL regions [Fig. S2(b-c)] at room temperature (RT). Here, tBL also features strong indirect band emission like nBL as smaller twist angles allow strong interlayer coupling [see Fig. S2(d)].
\\
\\
Next, we perform steady-state PL spectroscopy on D1 with 532 nm CW laser under dual gating. Fig. S4 shows the individual PL spectra. The color plot in Fig. 2(b) clearly depicts three different regimes of PL intensity variation with gating. Fig. 2(c) shows the intensity variation of different peaks as a function of $V_{g}$, extracted from the fitting of individual spectra (see Fig. S4). At low $V_{g}$ around 0 V, the PL intensity is highest around $X_{II}$ peak. The intensities of $X_{II}$, $X_{III}$ and $X_{IV}$ peaks quench rapidly as $V_{g}$ increases in both positive and negative directions. We observe that the peak positions of $X_{I}$-$X_{IV}$ are relatively unchanged with varying $V_{g}$. Another sample, D5, also exhibits weak variation of peak positions with gating (see Fig. S5). This suggests that the interlayer interaction remains independent of the doping density resulting in no net spatial modulation of the electronic band gap. Thus, the moir\'e potential of intralayer excitons remains similar at all levels of gating assuming a negligible effect of doping on the binding energy of the excitons. We therefore attribute the gating induced quenching of $X_{II}$-$X_{IV}$ peaks to the formation of their charged exciton complexes. As the intensity of $X_{I}$ peak does not change significantly with relatively small $V_{g}$, we rule out the trion formation from $X_{I}$ at low $V_{g}$. As a result, we attribute the additional peak ($T_{I}$) observed with gating in Fig. 2(b-c) to the trion formed from the delocalized higher energy excitons which is discussed in detail in the next section.
\\
\\
Next, we probe the localized and delocalized nature of the different moir\'e excitons in tBL-MoS$_2$ using power-dependent PL measurements with 531 nm pulsed excitation (see Methods). Fig. 2(d) represents the color plot of the PL spectra with varying optical power at $V_{g}$ = 0, which depicts a clear intensity variation of $X_{I}$, $X_{II}$ and $X_{III}$ peaks. The corresponding power law of the PL intensities is shown in Fig. 2(e). The saturation of PL intensity of $X_{I}$ peak owing to the moir\'e localization is evident from the experimental data which exhibits a power law exponent of $\sim 0.45$. In contrast, the $X_{II}$ and $X_{III}$ exhibit an almost linear power law corroborating the delocalized nature of higher energy moir\'e excitons.
\\
\\
Further, we perform time-resolved photoluminescence (TRPL) measurements at different $V_{g}$ in the spectral window of the intralayer moir\'e excitons to study their decay dynamics. The decay components are extracted using a bi-exponential fitting, $I_{PL}(t) = A_1\,e^{-t/\tau_{1}} + A_2\,e^{-t/\tau_{2}}$, of the TRPL data deconvoluted with the instrument response function (IRF) [see methods and Fig. S6]. The sub-10 ps range of $\tau_{1}$ is comparable to the $A_{1s}$ exciton in a 1L-MoS$_2$, while faster than the typical decay timescale of interlayer moir\'e excitons\cite{scuri2020}. We believe such an observation is due to a combination of (a) shallower moir\'e potential depth compared to interlayer excitons, and (b) enhanced Coulomb scattering induced nonradiative decay rate due to dispersion-less nature of the moir\'e exciton sub-bands. Another tBL-MoS$_2$ sample, D6 also shows similar results (see Fig. S7). The other decay component, $\tau_{2}$ exhibits a decay time of few 100s of ps. We note that the ratio ($A_{2}/A_{1}$) of the coefficients of $\tau_{2}$ to that of $\tau_{1}$ is about 4.5--fold higher in tBL-MoS$_2$ compared to 1L-MoS$_2$ at $V_{g}$ = 0 V. This indicates an increased contribution of slower decay time component in tBL-MoS$_2$. The relatively flat moir\'e exciton sub-band induced by the mSL favours more population of excitons outside the light cone compared to the 1L-MoS$_2$, as illustrated in Fig. 2(f).
\\
\\
\textbf{Gate-tunable moir\'e exciton-resonant Raman scattering:} Recent studies highlight emerging coupling mechanisms between the moir\'e excitons and phonons\cite{mcdonnell2021,shinokita2021} which can play an important role in the dynamics of moir\'e exciton ensembles. Using the tBL-MoS$_2$, here we demonstrate tunable resonant Raman scattering by means of electrically controlled oscillator strength of moir\'e excitons discussed in the previous section. tBL-MoS$_2$ exhibits discernable $E_{2g}^{1}$, $A_{1g}$ and $2LA$ peaks with 633 nm (1.96 eV) excitation [see Fig. S8(a)]. From linearly co-polarized Raman measurements with 633 nm at 4.5 K, we observe that the intensity of both $E_{2g}^{1}$ and $A_{1g}$ peaks decreases with an increase in the $V_{g}$ in either positive or negative direction [top panels of Fig. 3(a-b)]. At $V_{g}$ = 0 where the doping is very weak, tBL features multiple moir\'e sub-band emission among which $X_{III}$ is closely resonant to 633 nm excitation [Fig. 3(c)]. Such incident resonance manifests in strong resonant Raman scattering of all phonon modes. As $|V_{g}|$ increases, the oscillator strength of the $X_{III}$ exciton significantly reduces [Fig. 2(b-c)], which suppresses the excitation resonance, and hence  a subsequent reduction in the Raman intensity of all the phonon modes.
\\
\\
On the other hand, in an nBL-MoS$_2$ [Fig. S8(b)], only $A_{1g}$ shows a monotonically decreasing intensity with increasing $V_{g}$ while the $E_{2g}^{1}$ remains unchanged in intensity [bottom panels of Fig. 3(a-b)]. This trend correlates with the symmetry dependent renormalization of phonon modes under doping\cite{chakraborty2012,lu2017}. With electron doping, softening and broadening of the $A_{1g}$ occurs due to increased electron-phonon coupling (EPC) while $E_{2g}^{1}$ remains insensitive to doping from symmetry forbidden EPC. The comparison of the gate controlled resonant Raman scattering between tBL-MoS$_2$ and nBL-MoS$_2$ thus clearly distinguishes the tunability of moir\'e exciton-phonon scattering in a tBL-MoS$_2$, which can be useful to probe intriguing moir\'e physics.
\\
\\
\textbf{Emission from multiple intralayer moir\'e trions:} In contrast to 532 nm CW measurements [Fig. 2(b-c)] where only a single trion peak ($T_{I}$) emerges with gating, pulsed excitation of D1 with 531 nm laser shows evolution of two different peaks, $T_{I}$ and $T_{II}$ to the left of $X_{I}$ [see Fig. 4(a-b)]. The corresponding individual PL spectra are shown in Fig. S9 along with the representative spectral fitting. We consistently observe the multiple emission features of moir\'e exciton sub-bands under pulsed excitation as well. Note that, with pulsed laser, the peak positions of moir\'e excitons are slightly blue-shifted relative to that of CW laser, indicating a higher excitation density (see Fig. S9). The subsequent intensity quenching of the moir\'e exciton peaks (Fig. S9) with the evolution of both $T_{I}$ and $T_{II}$ [Fig. 4(b)] indicates that they are charged exciton complexes. Further, at higher $V_{g}$, $T_{II}$ exhibits a continuous increase in its intensity, while $T_{I}$ displays a conspicuous decrease in its intensity. Another device D7 (Fig. S10) also exhibits a similar trend at very high $V_{g}$ even under 532 nm CW excitation.
\\
\\
From the PL spectra [Fig. S2(d)], we infer that tBL-MoS$_2$ has an indirect band gap similar to the nBL-MoS$_2$. Hence, the electrons and holes injected through gating respectively occupy the CB minimum at the $\bs{\Lambda}$ point ($\Lambda_{c}$)\cite{rama2011,zhao2013} and the VB minimum at the $\bs{\Gamma}$ point ($\Gamma_{v}$). Thus, we attribute the intralayer negative (positive) trions to be formed by the binding of the additional electron (hole) from the $\Lambda_{c}$ ($\Gamma_{v}$) valley with the optically induced intralayer excitons at the $\bs{K}$ and $\bs{K^{\prime}}$ valley. Hence, to understand the experimental observation of multiple trion peaks and their intensity variation with $V_{g}$, we analyse the moir\'e potential profiles of CB and VB corresponding to both indirect ($\Lambda_c$ and $\Gamma_v$) and direct band gap minima ($K_c$ and $K_v$) through the solutions of Poisson equation, as shown in the left panel of Fig. 5(a) [see Note-II under Fig. S11 in SI for simulation details].
\\
\\
We find that the moir\'e potential across CB or VB minima is relatively unperturbed at lower gating. As the gating increases further, the CB (VB) spatial profiles get modulated such that the amplitude of electronic moir\'e potential at $\Lambda_c$ ($\Gamma_v$) decreases and that at $K_c$ and $K_v$ increases. At very high gating, the moir\'e potential at $\Lambda_c$ ($\Gamma_v$) flattens out completely leading to delocalized electrons (holes). On the other hand, the electrostatic potential at high gating amplifies [see Fig. S11] the fluctuation in the moir\'e potential at $K_c$ and $K_v$ such that the corresponding band profiles evolve in-phase in contrast to the initial out-of-phase condition. This renders an interesting trend of transfer of the moir\'e potential spatial modulation from the $\Lambda_c$ ($\Gamma_v$) to the $K_c$ and $K_v$ with positive (negative) gating. However, gating does not change the effective spatial modulation of the indirect and direct band gaps [see the right panel of Fig. 5(a)], and hence the moir\'e confinement remains similar for intralayer exctions at all $V_g$.
\\
\\
The modulation of the depth of moir\'e potential of electrons (holes) with positive (negative) gating coupled with different levels of optical excitation results in the formation of different trion species, which is illustrated in Fig. 5(b).
At lower positive gating where the moir\'e potential for electrons at $\Lambda_c$ [$V_{M}^{e}(\Lambda)$] is unperturbed, the injected carriers from the contact occupy the moir\'e traps where the mSL potential is minimum. In this regime, the electron density (\textit{n}) is smaller than the moir\'e trap density ($N_{m}$) and hence fewer moir\'e traps are occupied by electrons. At such low level of doping, coupled with low exciton density ($N_{ex}$) at low excitation power, it is less probable to populate an exciton into an electron pre-occupied moir\'e trap. Thus, the moir\'e localized excitons ($X_{L}$) and electrons ($e_{L}$) likely stay in different traps which inhibits them from binding together. On the other hand, delocalized excitons ($X_{D}$) get attracted to $e_{L}$ in the moir\'e traps to form a localized trion, $T_{DL}$. Even at higher $N_{ex}$, the presence of fewer $e_{L}$ with lower gating favours only $T_{DL}$.
\\
\\
With further increase in the positive gating, $V_{M}^{e}(\Lambda)$ reduces such that $e_{L}$ exist along with the delocalized electrons ($e_{D}$) when the thermal energy is comparable to the depth of the  moir\'e trap. At low $N_{ex}$, $X_{D}$ binds with both $e_{L}$ and $e_{D}$ to form localized ($T_{DL}$) and delocalized ($T_{DD}$) trions, respectively. In this regime, when \textit{n} $\sim\:N_{m}$, most of the moir\'e traps get occupied with electrons. Such high level of doping coupled with high $N_{ex}$ facilitates the creation of excitons in the electron occupied moir\'e traps resulting in the formation of a localized trion ($T_{LL}$) in addition to the delocalized trion ($T_{DD}$).
\\
\\
Further increase of gating leads to complete delocalization of electrons. Here, both $X_{L}$ and $X_{D}$ attract $e_{D}$ to form a localized trion, $T_{LD}$ and a delocalized trion, $T_{DD}$ respectively. Due to the relatively stable configuration of $T_{LD}$ from its lower energy, the formation of $T_{LD}$ is more favorable at higher $V_{g}$. Hence, we infer that the experimentally observed intensity trend [see Fig. 4(b)] of trion peaks, $T_{I}$ and $T_{II}$ at high $V_{g}$ correspond to the delocalized trion, $T_{DD}$ and the localized trion, $T_{LD}$, respectively. From the conceptual model of trion formation presented in Fig. 5(b), we correlate the different trions as a function of $V_{g}$, as shown in Fig. 4(b). Note that the larger exciton density facilitated by the pulsed laser excitation favors formation of additional trions from the localized excitons and thus enables the observation of $T_{II}$ in addition to $T_{I}$ with gating. The separation between $X_{I}$ and $T_{II}$ is about $\sim$5-7 meV higher than that between $X_{II}$ and $T_{I}$ indicating a stronger trion binding energy with the moir\'e trapped exciton. $X_{II}$-$T_{I}$ is about $\sim$30 meV (see Fig. S9), similar to the trion binding energy in 1L-MoS$_2$\cite{jadczak2021,golovynskyi2021}.
\\
\\
As the $T_{I}$ and $T_{II}$ peaks extracted from spectral fitting are quite broad (30-40 meV), it is experimentally difficult to differentiate their localized and delocalized phases at lower $V_{g}$. However, as $T_{LD}$ ($T_{II}$) is dominant and $T_{DD}$ ($T_{I}$) is suppressed at very high $V_{g}$, it is possible to segregate the localized nature of $T_{LD}$ using power-dependent PL measurements. Fig. 4(c) shows such a power-dependent PL with 531 nm pulsed laser at $V_{g}$ = 5 V, where PL intensity shows a saturation trend with a power law exponent of $\sim 0.53$. This clearly indicates the localized nature of trions at high gating arising from the saturation of moir\'e localized excitons at higher optical power. We also perform TRPL measurements in the spectral window covering both $T_{I}$ and $T_{II}$ whose dominant decay component ($\tau_{1}$) is shown in Fig. 4(d) (see Fig. S12 for the corresponding bi-exponential deconvolution of TRPL data). The faster decay with increasing $|V_g|$ likely results from a stronger Coulomb scattering between the trions and the increasingly delocalized carriers.
\\
\\
\textbf{Gate-tunable valley coherence of the moir\'e excitons:}
The intralayer moir\'e excitons inherit the valley characters of the individual layers, and thus exhibit strong valley coherence that can be initialized by the excitation through a linearly polarized light. The valley coherence is estimated by the degree of linear polarization (DOLP) of the output as measured by DOLP=$\frac{I_{co}-I_{cross}}{I_{co}+I_{cross}}$. Here, $I_{co}$ and $I_{cross}$ are the measured PL intensity in the co- and cross- polarization mode with respect to the incident linear polarization. In Fig. 6(a), we show the measured DOLP (in symbols) for the localized exciton ($X_I$) as a function of $V_g$ under near-resonant (633 nm) excitation, which exhibits a striking W-shape. The corresponding PL spectra are shown in Fig. S13(a). In particular, both in \textit{n}-type and \textit{p}-type doping regimes, the DOLP first decreases with $V_g$, followed by an increase.
\\
\\
In order to explain the trend, we model the DOLP using Maialle-Silva-Sham (MSS) mechanism \cite{MSS1993, garima2021}, as indicated by the solid trace in Fig. 6(a). The decoherence mechanism is schematically explained in Fig. 6(b-c). Under near-resonant excitation, the exciton is generated in the light cone of the near-flat moir\'e sub-band with a pseudo-spin vector $(S_x,S_y,S_z)=(1,0,0)$. The localized $X_I$ excitons encounter Coulomb scattering with an increasing density of delocalized electrons (holes) due to an increase in the positive (negative) $V_g$. Note that, unlike free excitons with parabolic bands, the flat-band nature of $X_I$ allows for a change in the magnitude of the center of mass momentum ($\mathbf{Q}$) even under elastic scattering [Fig. 6(b)]. The resulting change in $\mathbf{Q}$ of the exciton causes a non-zero angle between the pseudo-spin vector and the spin-exchange induced pseudo-magnetic field. This in turn results in a precession of the pseudo-spin vector [Fig. 6(c)]. The random pseudo-spin precession accumulates through multiple scattering events till the radiative recombination of the exciton. The steady-state DOLP is obtained by averaging $S_x$ over the entire $\mathbf{Q}$ space, and is given by \cite{garima2021}
\begin{equation}
\mathbf{G} = \frac{\mathbf{S}(\mathbf{Q})}{\tau} - \mathbf{\Omega}(\mathbf{Q}) \times \mathbf{S}(\mathbf{Q}) - W\sum_{\mathbf{Q^\prime}}[\mathbf{S}(\mathbf{Q^\prime})-\mathbf{S}(\mathbf{Q})]
\end{equation}
Here $\mathbf{G}=[1\: 0\: 0]$ is the generation rate vector, $\tau$ is the lifetime of the exciton as obtained from TRPL measurements, and $\mathbf{\Omega}$ is the precession frequency. The fitted scattering rate ($W$) that matches the experimental data is shown in Fig. 6(d), indicating an exponential dependence in $V_g$. The variation changes to a polynomial behavior at high positive $V_g$. This is in good agreement with the fact that the Coulomb scattering generated from the  $V_g$-dependent delocalization of the carriers results in such DOLP dependence.
\\
\\
While the initial degradation in the observed DOLP with $V_g$ at small $|V_g|$ is due to the enhanced scattering rate, when the scattering rate crosses some threshold, the DOLP starts improving. This is due to the motional narrowing effect\cite{garima2021}, where the scattering rate becomes faster than the precession frequency. Under such a situation, due to random phase cancellation, the net accumulated random phase in the pseudo-spin vector becomes small, restoring the high DOLP value. A similar trend is also observed for the delocalized exciton ($X_{II}$) due to the increasing scattering rate with the delocalized carriers [see Fig. S13(b)].

\section*{Conclusion}
To summarize, the following are the key observations from our experiments on the MoS$_2$ twisted bilayer samples with a near-$0^0$ twist angle in H-type stacking.
\begin{enumerate}
  \item We observe the emission from multiple intralayer excitons induced by the moir\'e sub-bands.
  \item We demonstrate the electrical tunability of these intralayer moir\'e exciton emission through the formation of intralayer moir\'e trions.
  \item We identify the localized and delocalized phases of moir\'e exctions and trions by varying the photon flux and carrier concentration. While the spatial modulation of excitonic bandgap remains unaltered with gating, we find that the variation in moir\'e potential for individual charge carriers (electrons and holes) tunes their corresponding localization and thereby modulates the phase of the moir\'e trion, which can be further controlled by the optical excitation density. Interestingly, the electrostatic modulation of moir\'e potential of carriers at the indirect band gap minima results in a transfer of the moir\'e potential fluctuation to the corresponding direct band gap minima offering ways to control the localization of carriers at different momenta.
 \item We also demonstrate the gate tunable valley coherence and exciton-phonon coupling strength for the moir\'e excitons.
\end{enumerate}

Such electrical control of the moir\'e excitons and trions exemplifies the possible tunability of quantum phenomena in twisted structures of layered materials, and points to the possibilities of engineering photonic and valleytronic devices with moir\'e superlattices.

\section*{Methods}
\subsection*{Device fabrication}
We prepare the tBL samples at very small $\theta$ (ranging from $0^0$ to $3^0$) in H-type stacking using `tear and stack' dry transfer method. First, a hBN flake is transferred on to a pre-patterned Au line on a SiO$_2$/Si substrate. This is followed by a few-layer graphene (FLG) transfer on hBN/SiO$_2$/Si region touching the second Au line. Next, a monolayer (1L) MoS$_2$ is exfoliated on to another SiO$_2$/Si substrate and then a PDMS stamp exfoliated with hBN is aligned with the substrate to pick up a portion of 1L-MoS$_2$ tearing it at the edge of hBN. The substrate is then rotated by $180^0-\theta$ using a rotation stage. The rotation of $\theta$ is also confirmed by measuring the twist angle of pre-defined alignment markers on the substrate using high resolution microscope imaging. After the desired adjustment of $\theta$, the PDMS stamp with hBN and pre-picked 1L-MoS$_2$ is aligned and contacted with the substrate. The air-release is controlled carefully to ensure only a partial coverage of hBN region and the PDMS stamp is slowly withdrawn away from the substrate for a successful pick up of the remaining portion of 1L-MoS$_2$. This hBN with tBL-MoS$_2$ stack is transferred on to the hBN and FLG pre-transferred substrate such that one of the 1L-MoS$_2$ makes a contact with pre-transferred FLG. Finally, another FLG layer is transferred on top of this stack covering the tBL region completely, touching the third Au line. To improve the coupling between the constituent monolayers, the device is annealed in vacuum ($\sim 10^{-6}$ torr) at $200^{0}$ C for 5 hours. Au and FLG serve as the bottom and the top gate electrodes, respectively, while the FLG contact touching the tBL-MoS$_2$ is connected to the ground.
\subsection*{PL spectroscopy under gating}
All the photoluminescence (PL) measurements on the samples are carried out in a closed-cycle cryostat at $4.5$ K using a $\times50$ objective (0.5 numerical aperture) in reflection geometry. First, the top and bottom gate voltages are applied using Keitheley 2636B sourcemeter and then the PL spectra are collected using a spectrometer with 1800 lines per mm grating and CCD. During all the gated-PL measurements discussed in this work, the gate leakage current was considerably low (< 1 nA) as shown in Fig. S4. $532$ and $633$ nm CW excitations are used at $50$ and $85$ $\mu$W of laser power respectively for exposure time of $10$ or $20$ s. $531$ nm pulsed laser is operated at a repetition rate of $10$ MHz for PL characterization. Power-dependent PL spectra from tBL-MoS$_2$ devices at a required gate voltage are obtained by using a $531$ nm pulsed laser at a repetition rate of $80$ MHz. The power calibration is performed using a power meter (LaserCheck$^{TM}$) at the sample location in the spectrometer setup.
\subsection*{TRPL}
Our TRPL setup comprises of a $531$ nm pulsed laser head (LDH-D-TA-530B from PicoQuant) controlled by the PDL-800 D driver, a photon counting detector (SPD-050-CTC from Micro Photon Devices), and a Time-Correlated Single Photon Counting (TCSPC) system (PicoHarp 300 from PicoQuant). The pulse width of the laser is $\sim 40$ ps with a maximum repetition rate of $80$ MHz. The decay dynamics of moir\'e excitons (trions) is measured by using a $635 (647.1)$ nm bandpass filter (FWHM-$10$ nm) coupled with a $600$ nm longpass filter. The instrument response function (IRF) has a decay of $\sim 23$ ps and an FWHM of $\sim 52$ ps. The deconvolution of the TRPL data with the IRF is performed in the QuCoa software from PicoQuant using a bi-exponential decay function. The deconvolution allows us to accurately extract down to $10\%$ of the IRF width\cite{becker2005}. During the TRPL measurements, $V_{tg}$ and $V_{bg}$ voltages are applied in equal magnitude. The average power of the laser used during TRPL measurements is 17 $\mu$W.

\begin{suppinfo}
The Supporting Information is available free of charge at XXXXX.
\begin{itemize}
    \item Mini Brillouin zone of moir\'e superlattice (S1); Room temperature characterization of devices-D1 and D2 (S2); Comparison of PL spectra of tBL-MoS$_2$ with different twist angles (S3); PL characterization of tBL-MoS$_2$ devices-D1 (S4), D5 (S5), D6 (S7) and D7 (S10); Pulsed laser excitation of tBL-MoS$_2$ device-D1 (S9); Electrostatics of tBL-MoS$_2$ from 2D Poisson simulation (S11); TRPL and decay analysis of excitons (S6) and trions (S12) in tBL-MoS$_2$ from D1; 633 nm linearly polarized Raman (S8) and PL (S13) from tBL-MoS$_2$ of D1.
\end{itemize}
\end{suppinfo}
\begin{acknowledgement}
This work was supported in part by a Core Research Grant from the Science and Engineering Research Board (SERB) under Department of Science and Technology (DST), a grant from Indian Space Research Organization (ISRO), a grant from MHRD under STARS, and a grant from MHRD, MeitY and DST Nano Mission through NNetRA. K.W. and T.T. acknowledge support from the Elemental Strategy Initiative conducted by the MEXT, Japan (Grant Number JPMXP0112101001) and  JSPS KAKENHI (Grant Numbers 19H05790, 20H00354 and 21H05233).
\end{acknowledgement}
\section*{AUTHOR CONTRIBUTIONS}
M.D. and K.M. designed the experiment. M.D. fabricated the devices, and performed the measurements. G.G and K.M. performed the time-resolved photoluminescence measurements. P.D. performed the electrostatic simulation. G.G. performed the modeling and analysis of valley coherence with inputs from K.M. K.W. and T.T. grew the hBN crystals. M.D. and K.M. performed the data analysis. M.D. and K.M. wrote the manuscript with inputs from others.
\section*{NOTES}
The authors declare no competing financial or non-financial interest.
\section*{Data Availability}
The data is available on reasonable request from the corresponding author.

\bibliography{refv1}

\newpage
\begin{figure*}[!hbt]
	\centering
	\includegraphics[scale=0.35] {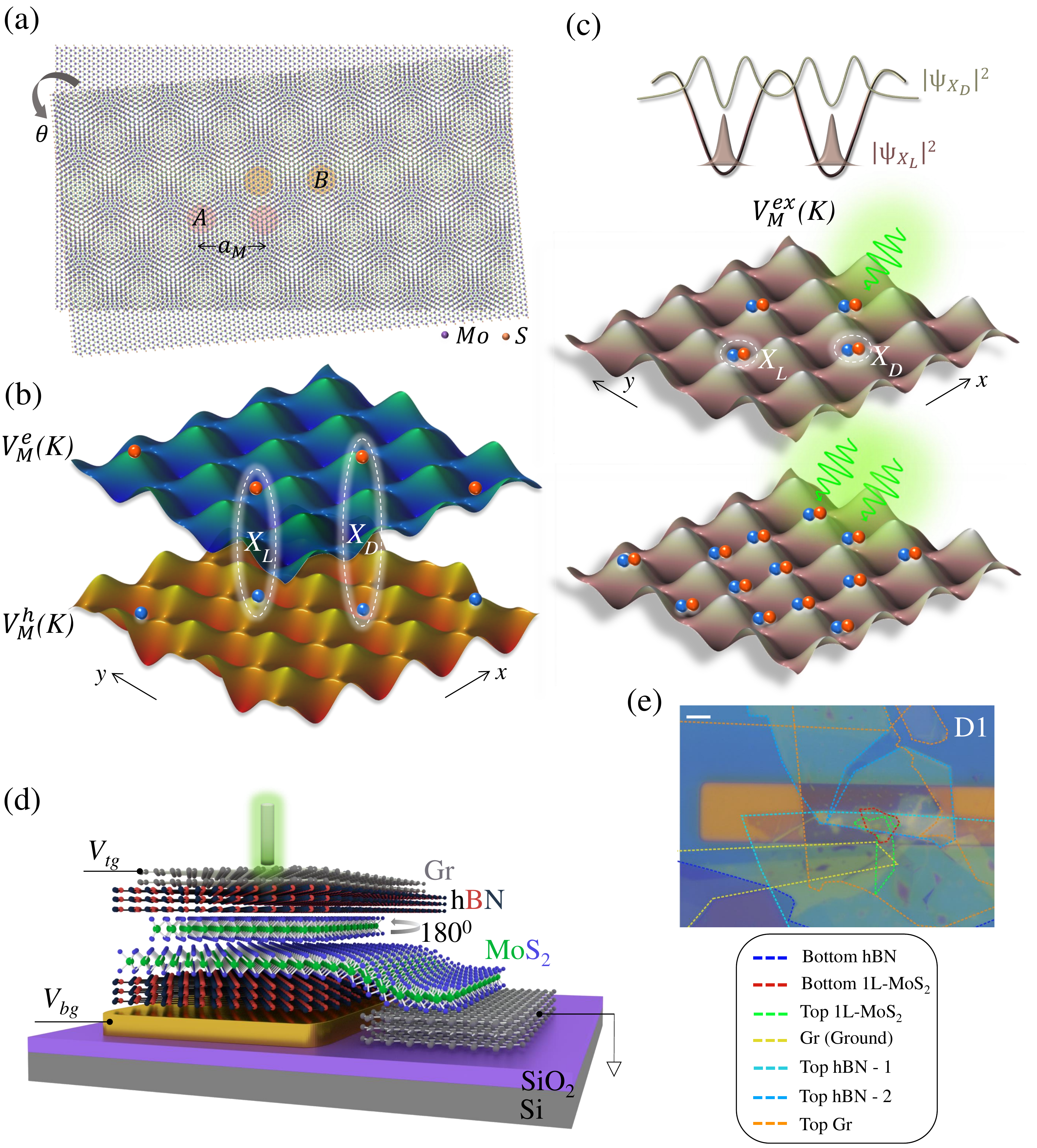}
	\caption{\textbf{Moir\'e superlattice of a twisted bilayer of MoS$_2$}. (a) Schematic of moir\'e superlattice (mSL) of twisted bilayer (tBL) of MoS$_2$ formed by H-type stacking with a twist angle, $\theta$. $a_{M}$ denotes the periodicity of mSL. A and B denote the regions with $H^{M}_{X}$ and $H^{M}_{M}$ stacking respectively.  (b) Schematic illustration of moir\'e potential of electrons and holes at $\bs{K}$ point (using eq. 1 noted in SI).  Red and blue spheres denote the optically generated electrons and holes, respectively, which bind together to form excitons. (c) Top panel: The localized and delocalized wavefunction distributions of the lowest energy exciton ($X_{L}$) and the higher energy excitons ($X_{D}$) respectively. Middle and bottom panels illustrate the moir\'e potential of intralayer excitons ($V^{ex}_{M}$) and the corresponding low and high intensity optical excitation of tBL respectively. (d) Schematic of tBL-MoS$_2$ dual gated device. (e) Optical image of tBL-MoS$_2$ with near-$0^{0}$ H-type stacking (D1). Scale bar is 5 $\mu$m. The individual layers are highlighted with colored dashed lines.}\label{fig:F1}
\end{figure*}
\pagebreak

\begin{figure*}[!hbt]
	\centering
	\includegraphics[scale=0.45] {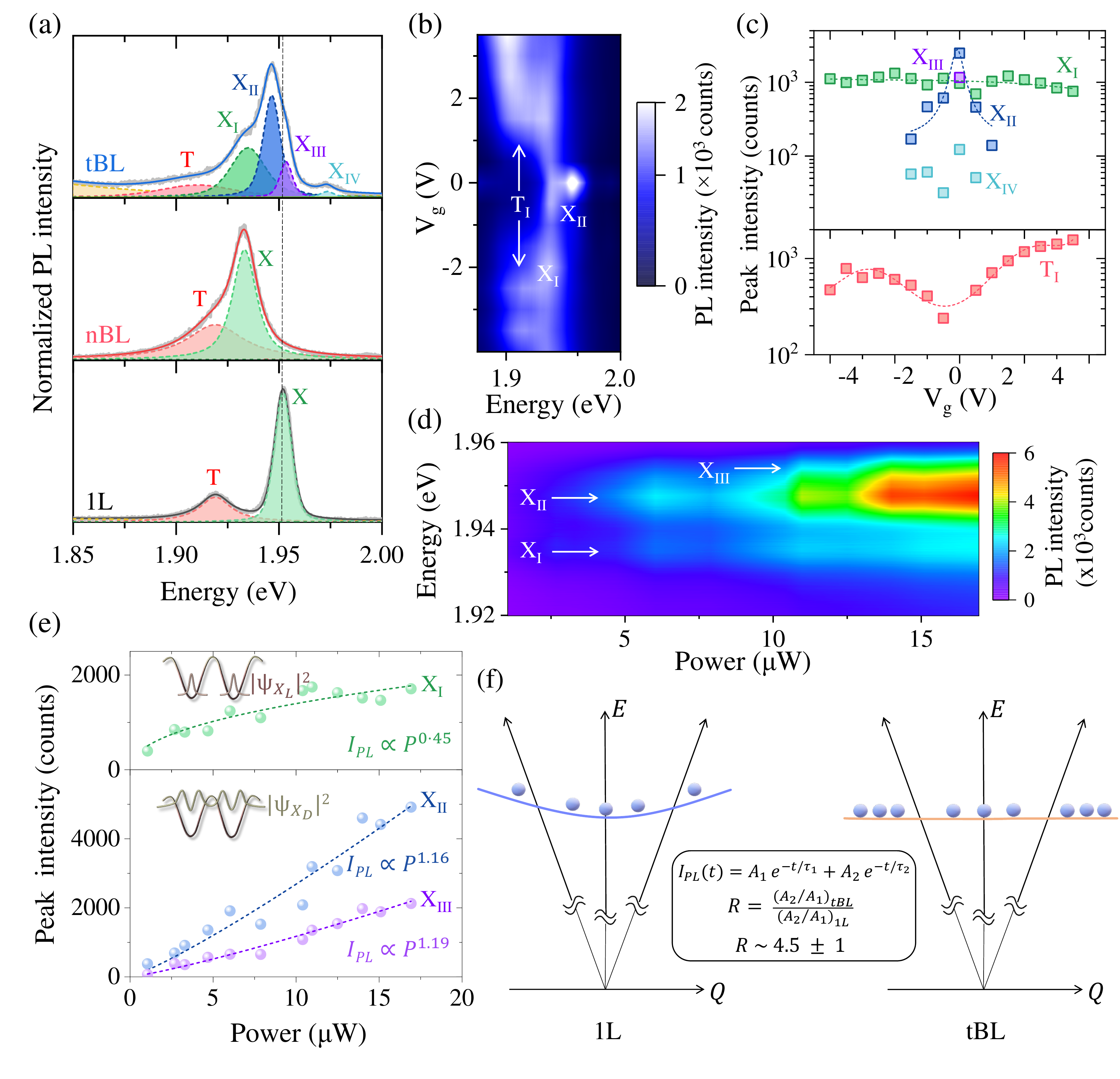}
	\caption{\textbf{Moir\'e excitons in a twisted bilayer of MoS$_2$}. (a) PL spectra ($4.5$ K) of twisted bilayer (tBL), natural bilayer (nBL) and monolayer of  MoS$_2$ with floating gates. Colored regions under the dashed lines represent the peaks obtained from the spectral fitting (solid colored lines) of the corresponding experimental spectra (grey lines). X and T denote the exciton and trion peaks respectively. Black dashed line denotes the position of $A_{1s}$ exciton of 1L-MoS$_2$. (b) Color plot of PL spectra (with 532 nm CW laser) of tBL-MoS$_2$ with varying gate voltage ($V_{g}$). (c) Peak intensity as a function of $V_{g}$ for moir\'e exciton (top panel) and trion (bottom panel) peaks. The dashes lines are a guide to the eye. Contd...}\label{fig:F2}
\end{figure*}
\pagebreak
\addtocounter{figure}{-1}
\begin{figure} [t!]
  \caption{(Previous page.) \textbf{Moir\'e excitons in a twisted bilayer of MoS$_2$}. (d) Color plot of PL spectra of tBL-MoS$_2$ with varying excitation power of 531 nm pulsed laser at $V_{g}$ = 0 V. (e) Power law of the PL intensity of moir\'e exciton peaks. The dashed lines highlight the corresponding fitting used to extract the power law exponent. $X_{I}$ (top panel) exhibits a sublinear power law confirming its localized nature, while $X_{II}$ and $X_{III}$ (bottom panel) exhibit an almost linear power law owing to their delocalized nature.  (f) Light cone and the corresponding exciton density in 1L-MoS$_2$ (left panel) and tBL-MoS$_2$ (right panel). The relatively flat moir\'e exciton sub-band induced by the mSL favours more population of excitons outside the light cone compared to the 1L-MoS$_2$, strengthening the slower decay component in tBL-MoS$_2$.}
\end{figure}

\begin{figure*}[!hbt]
	\centering
	\includegraphics[scale=0.45] {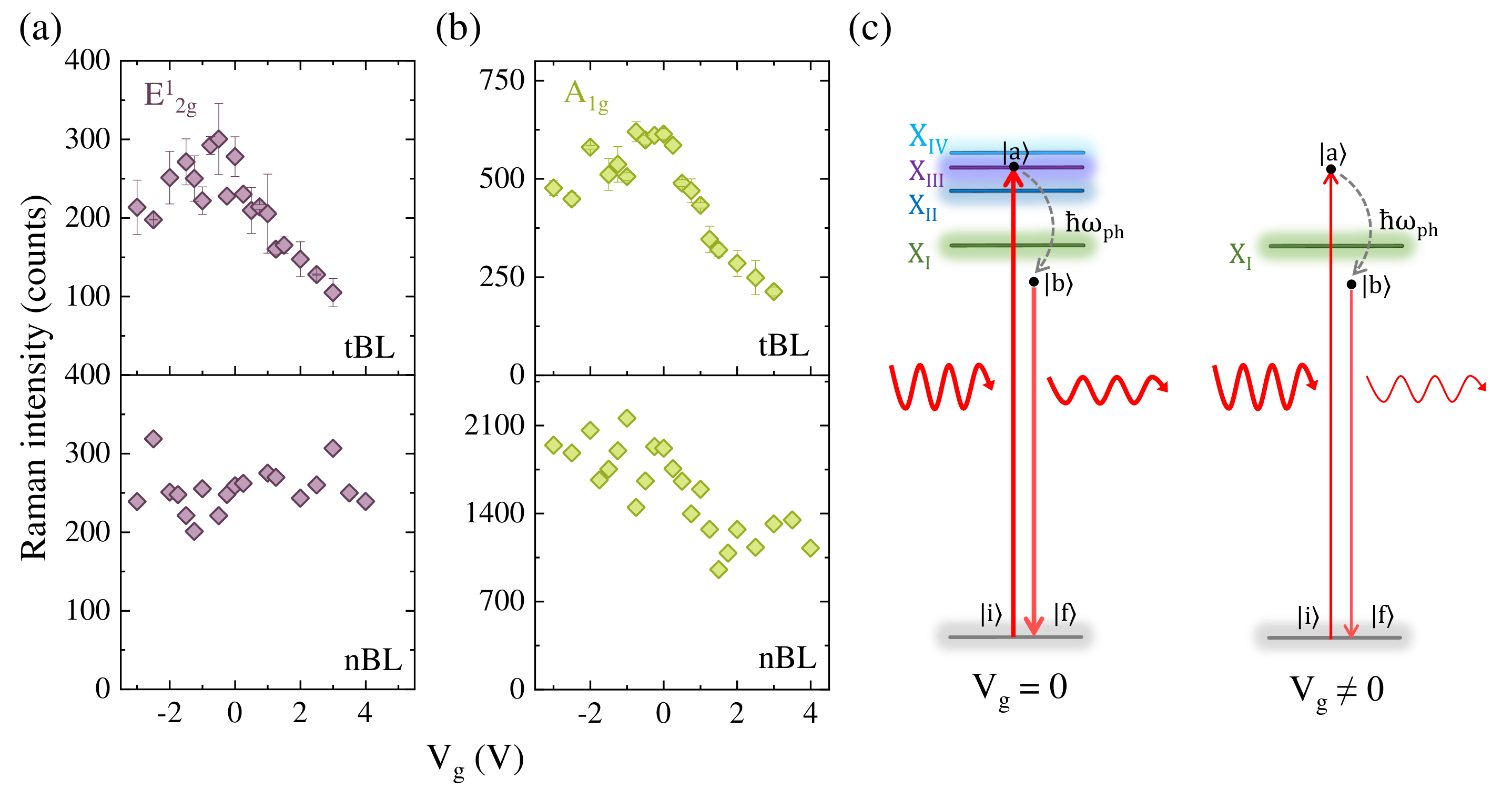}
	\caption{\textbf{Gate-tunable resonant Raman scattering through moir\'e exciton}. (a-b) Comparison of  $V_{g}$ dependence of Raman intensity of $E_{2g}^{1}$ and $A_{1g}$ modes between tBL (top panel) and nBL (bottom panel) of MoS$_2$. The peak intensities are extracted from spectral fitting of 633 nm linearly co-polarized Raman spectra. (c) Schematic of moir\'e exciton-resonant Raman scattering process under different gating conditions. A reduction in the oscillator strength of the moir\'e exciton resonant to 633 nm excitation causes a corresponding quenching in the Raman scattering process. $\ket{i}$ and $\ket{f}$ denote the initial and final states during the Raman scattering process while $\ket{a}$ and $\ket{b}$ denote the intermediate states.}\label{fig:F3}
\end{figure*}
\pagebreak

\begin{figure*}[!hbt]
	\centering
	\includegraphics[scale=0.6] {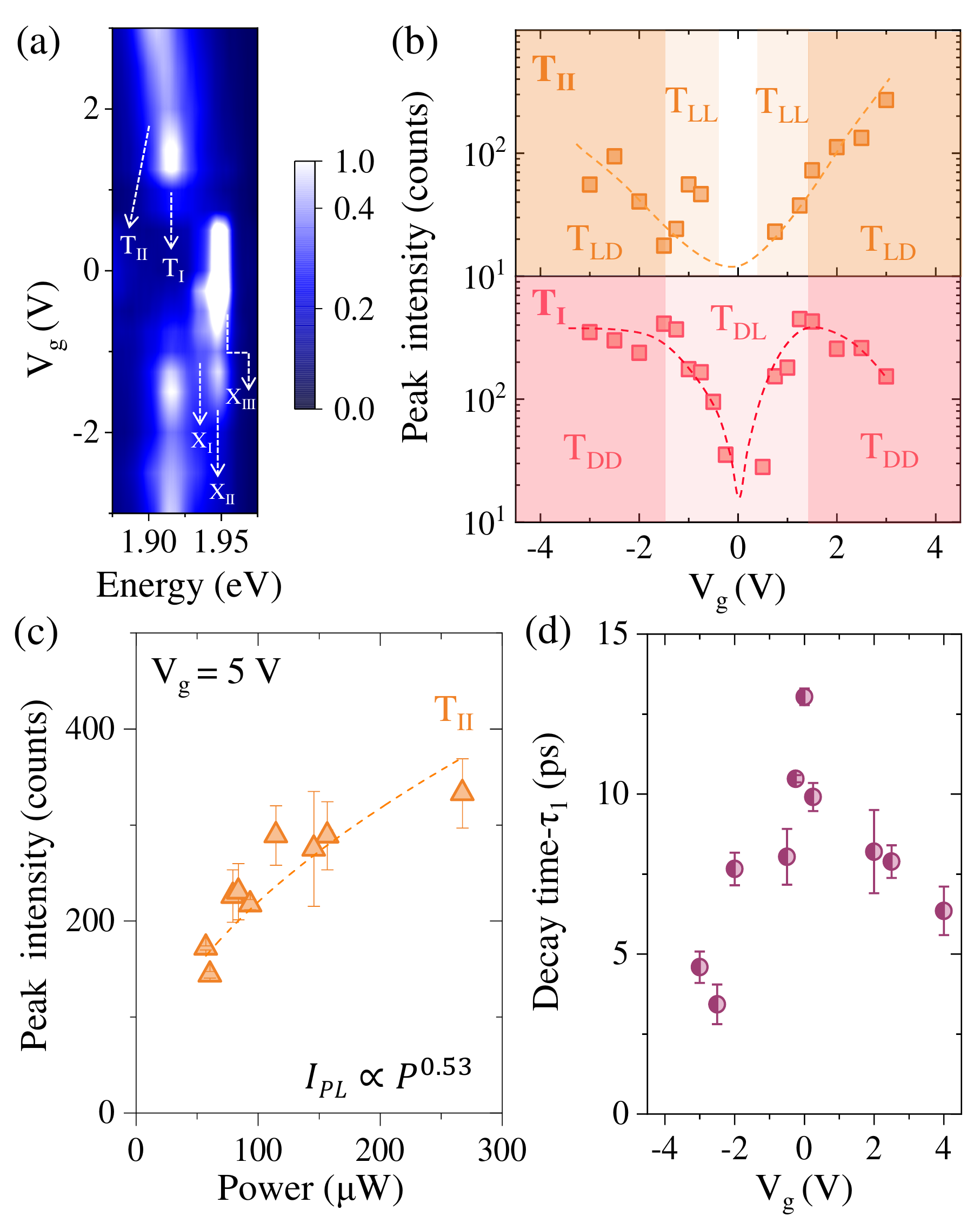}
	\caption{\textbf{Moir\'e trions in a twisted bilayer of MoS$_2$}. (a) Color plot of PL spectra of tBL-MoS$_2$ with varying $V_{g}$ from 531 nm pulsed laser. $T_{I}$ and $T_{II}$ denote the two trion peaks observed with gating. (b) Peak intensity of $T_{II}$ (top panel) and $T_{I}$ (bottom panel) as a function of $V_{g}$ extracted from the fitting of corresponding PL spectra. The dashed lines are a guide to the eye. (c) Power law of the trion peak, $T_{II}$ obtained at high $V_{g}$ = 5 V. This shows its localized nature owing to the formation of $T_{LD}$ kind of trion illustrated in Fig. 5(b). Dashed lines denote the corresponding fitting used to extract the power law exponent. (d) Decay time ($\tau_{1}$) as a function of $V_{g}$ extracted from TRPL measurements in the spectral window of moir\'e trions.}\label{fig:F4}
\end{figure*}
\pagebreak

\begin{figure*}[!hbt]
	\centering
	\includegraphics[scale=0.5] {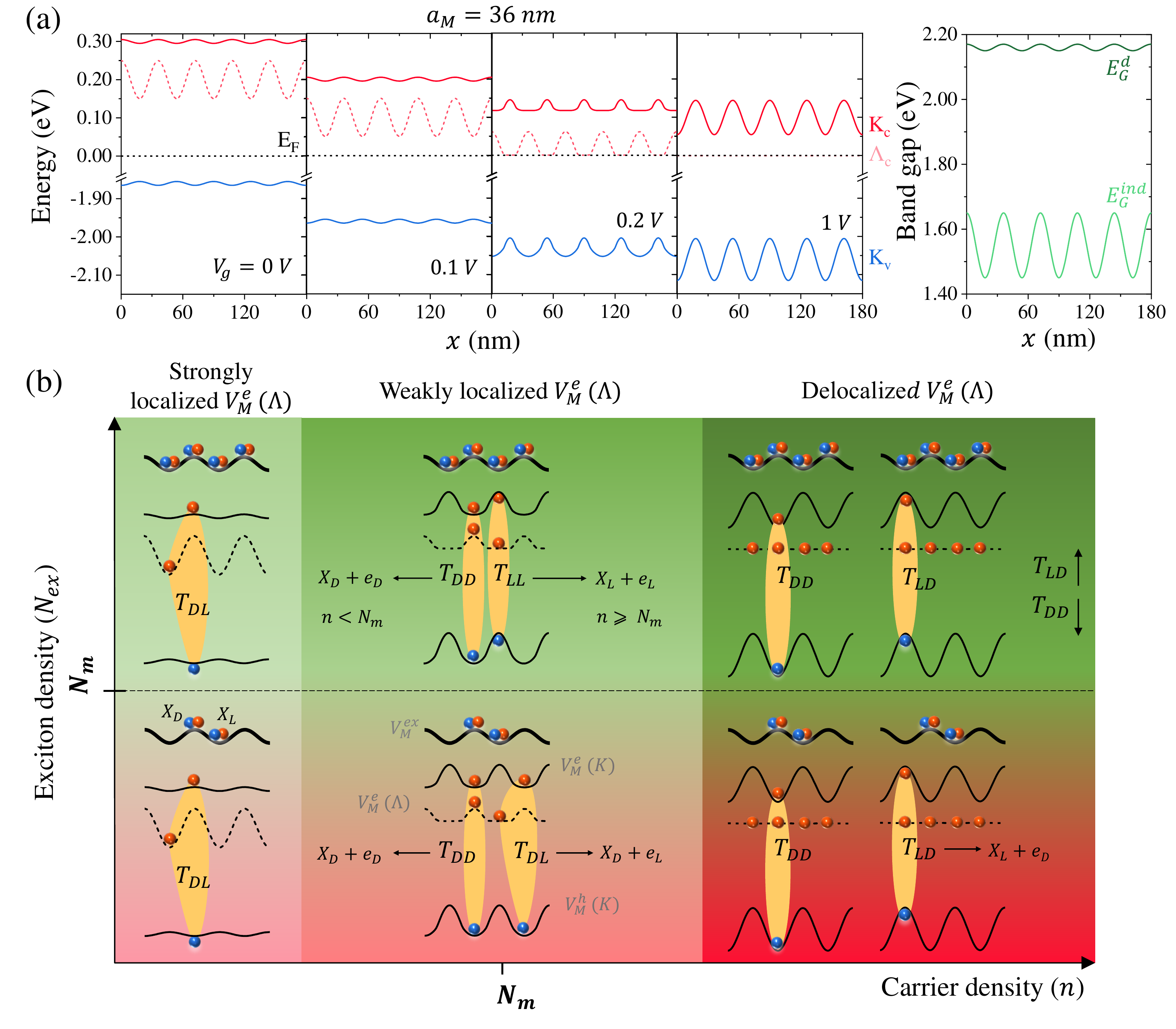}
	\caption{\textbf{Electrical tunability of moir\'e potential and trion formation}. (a) Left panel - Simulated spatial profiles of conduction (red) and valence (blue) bands at different $V_{g}$, obtained from the solution of Poisson equation. With an increase in positive $V_g$, the electronic moir\'e potential fluctuation is transferred from one valley to another, modulating the moir\'e localization of the corresponding carrier. Right panel - Spatial profiles of direct and indirect electronic band gap which remain unchanged (and, hence excitonic moir\'e confinement also remains unchanged) with $V_{g}$. Here, $a_{M}$ = 36 nm which corresponds to a twist angle of $0.5^0$. (b) Phase diagram of moir\'e trion formation as a function of exciton density ($N_{ex}$) and electron density (n) under positive gating. Here, different trions are denoted by $T_{AB}$ where the subscripts A and B denote the phase of the exciton and the electron respectively which are involved in the trion formation. The different phases of the trions shown in the model are associated to the experimental data in Fig. 4(b) accordingly.}\label{fig:F5}
\end{figure*}
\pagebreak

\begin{figure*}[!hbt]
	\centering
	\includegraphics[scale=0.5] {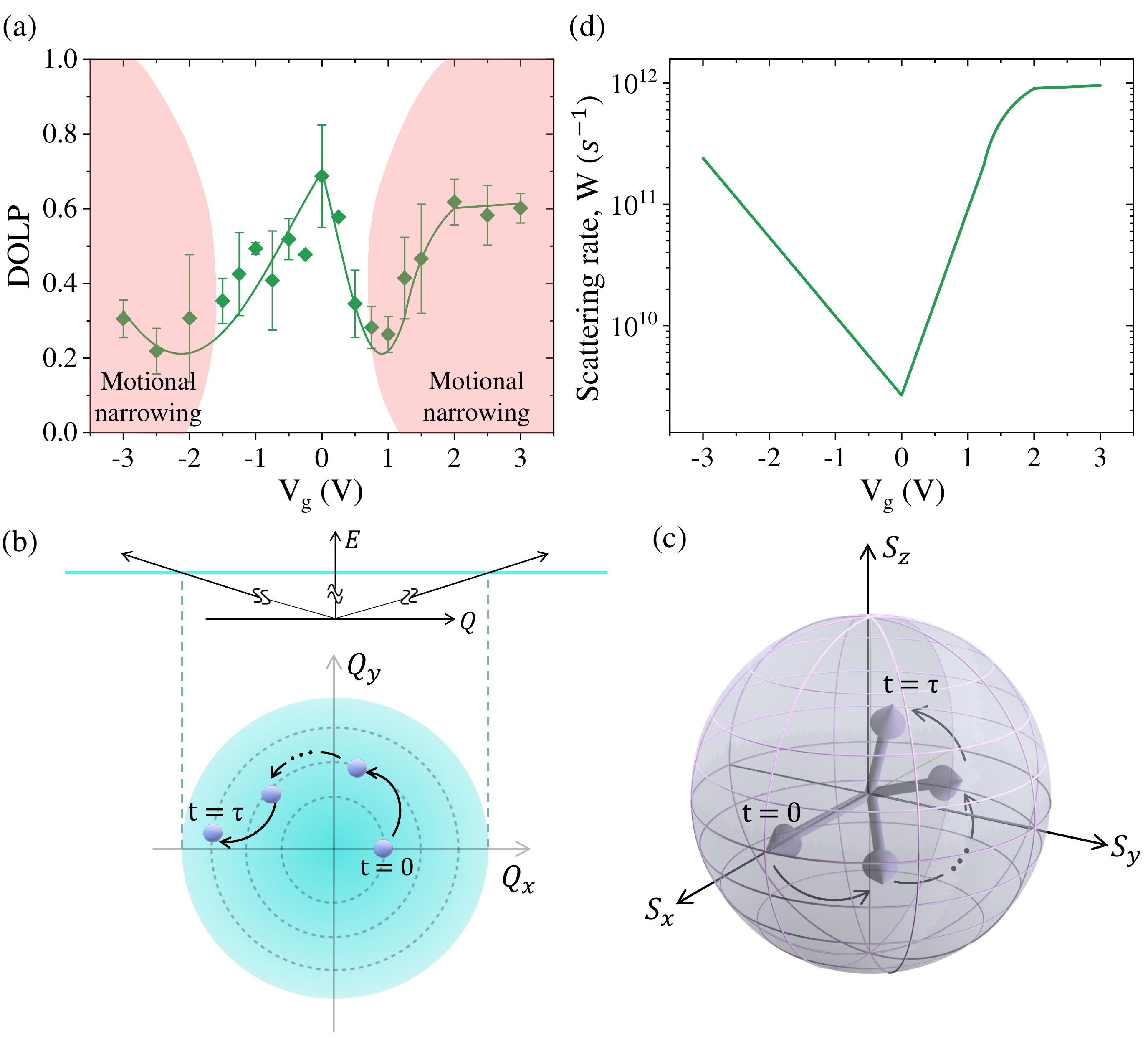}
	\caption{\textbf{Gate-tunable valley coherence of moir\'e excitons}. (a) Experimental trend (symbols) of degree of linear polarization (DOLP) of the localized exciton ($X_{I}$) with varying $V_{g}$ from D1 under 633 nm excitation. The solid trace in green denotes the fit to DOLP from the model using Maialle-Silva-Sham (MSS) mechanism. (b) Top panel: Flat moir\'e band along with the light cone. Bottom panel: Elastic Coulomb scattering of the moir\'e exciton within the light cone, with allowed change in COM momentum. (c) Evolution of the corresponding pseudo-spin vector of the moir\'e exciton in the Bloch sphere. (d) Fitted scattering rate ($W$) in logarithmic scale that corresponds to the solid trace in (a).}\label{fig:F6}
\end{figure*}
\pagebreak


\section*{Supplementary material}

# Note-I

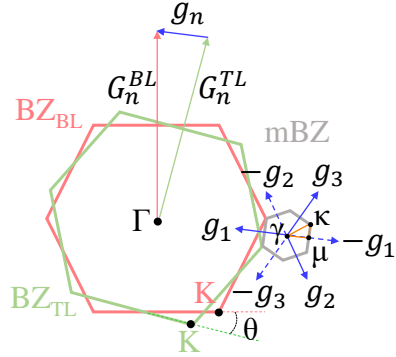

**Figure S1: Mini Brillouin zone of moiré superlattice**. Brillouin zones (BZ) of the top (TL) and bottom (BL) monolayers of MoS$_2$ in a $\theta$-tBL where $\boldsymbol{G_n^{TL}}$ and $\boldsymbol{G_n^{BL}}$ represent their corresponding basis of reciprocal lattice vectors. Bragg scattering in mSL leads to band-folding forming a mini BZ (mBZ) where $\boldsymbol{g_n} = \boldsymbol{G_n^{TL}} - \boldsymbol{G_n^{BL}}$ forms its basis of reciprocal lattice vectors.

Fig. S1 represents the Brillouin zone (BZ) of the top (TL) and bottom (BL) monolayers of MoS$_2$ with $\boldsymbol{G_n^{TL}}$ and $\boldsymbol{G_n^{BL}}$ as their respective reciprocal lattice vectors where $n$= 1, 2,..6. From a Fourier transformation, any perturbation over a physical quantity in the mSL, which is a function of in-plane co-ordinate, $\boldsymbol{r}$ can be expressed as a sum series of $e^{i(\boldsymbol{G_n^{TL}} - \boldsymbol{G_n^{BL}}) \cdot \boldsymbol{r}}$ terms.[1] Ignoring the fast oscillating components compared to $a_M$, the slowly oscillating term, $(\boldsymbol{G_n^{TL}} - \boldsymbol{G_n^{BL}})$ governs the moiré periodic modulation of electronic band structure. $\boldsymbol{g_n} = \boldsymbol{G_n^{TL}} - \boldsymbol{G_n^{BL}}$ corresponds to the reciprocal lattice vectors of mSL forming the basis for a mini brillouin zone (mBZ). Bragg reflection of the carriers from the periodic moiré potential leads to a folding of the original electronic bands into a series of closely spaced sub-bands in the mBZ. From a simple harmonic potential approximation in mSL for small $\theta$, the moiré potential ($V_M$) for electrons or holes in real space is expressed as[2,3]

$$V_M(\boldsymbol{r}) = \sum_{n=1}^{3} v_0 e^{i\boldsymbol{g_n} \cdot \boldsymbol{r}} + v_0^* e^{-i\boldsymbol{g_n} \cdot \boldsymbol{r}} \qquad (1)$$

where $v_0$ is a band-dependent complex parameter which varies with the interlayer translation and $\theta$.

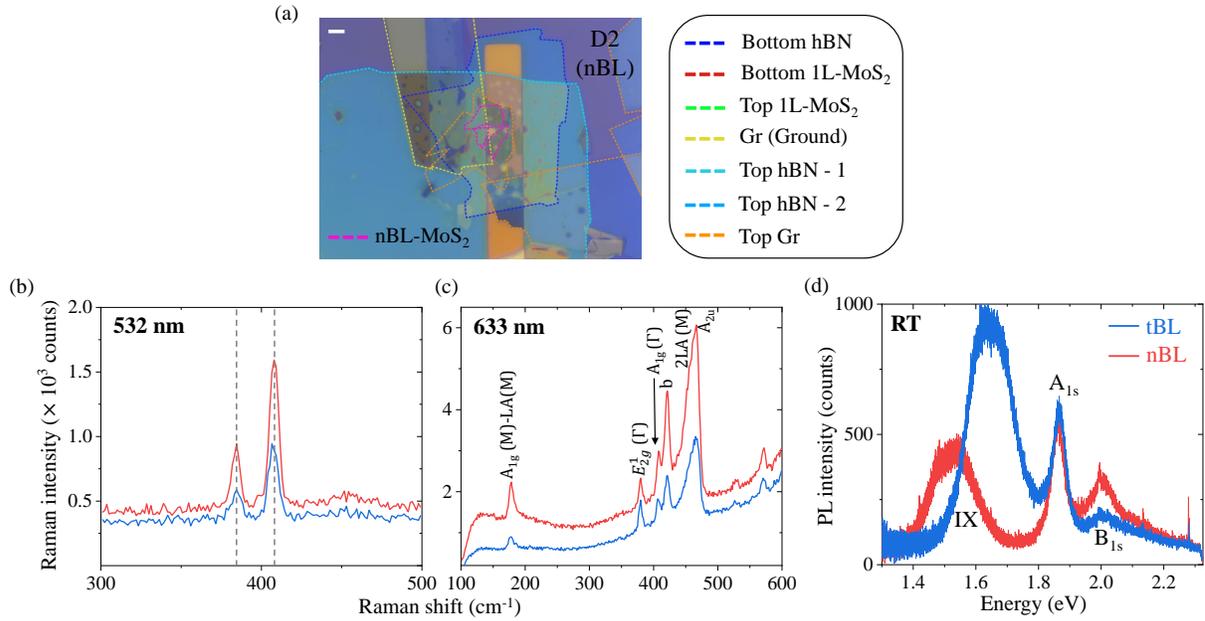

**Figure S2: Room temperature characterization of devices-D1 and D2.** (a) Optical image of D2 (nBL-MoS$_2$). Scale bar is 5 $\mu$m. The individual layers are highlighted with colored dashed lines. (b-c) Raman spectra of tBL- and nBL-MoS$_2$ regions with 532 nm and 633 nm excitation respectively at room temperature (RT). (d) PL spectra of tBL- and nBL-MoS$_2$ regions with 532 nm CW laser excitation at RT. IX indicates the indirect band emission.

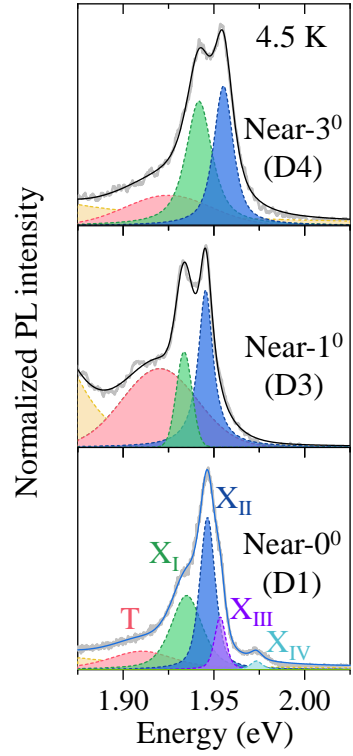

**Figure S3: Comparison of PL spectra of tBL-MoS$_2$ with different twist angles**. 532 nm PL spectra of tBL-MoS$_2$ with near-$3^0$ (D4), $1^0$ (D3) and $0^0$ (D1) twist angle ($\theta$) at 4.5 K under $V_g = 0V$. Colored regions under the dashed lines represent the peaks obtained from the spectral fitting (black lines) of the corresponding experimental spectra (grey lines). $X_I$ and $X_{II}$ exhibit a blue-shift and the higher energy peaks, $X_{III}$ and $X_{IV}$ disappear with increasing $\theta$.

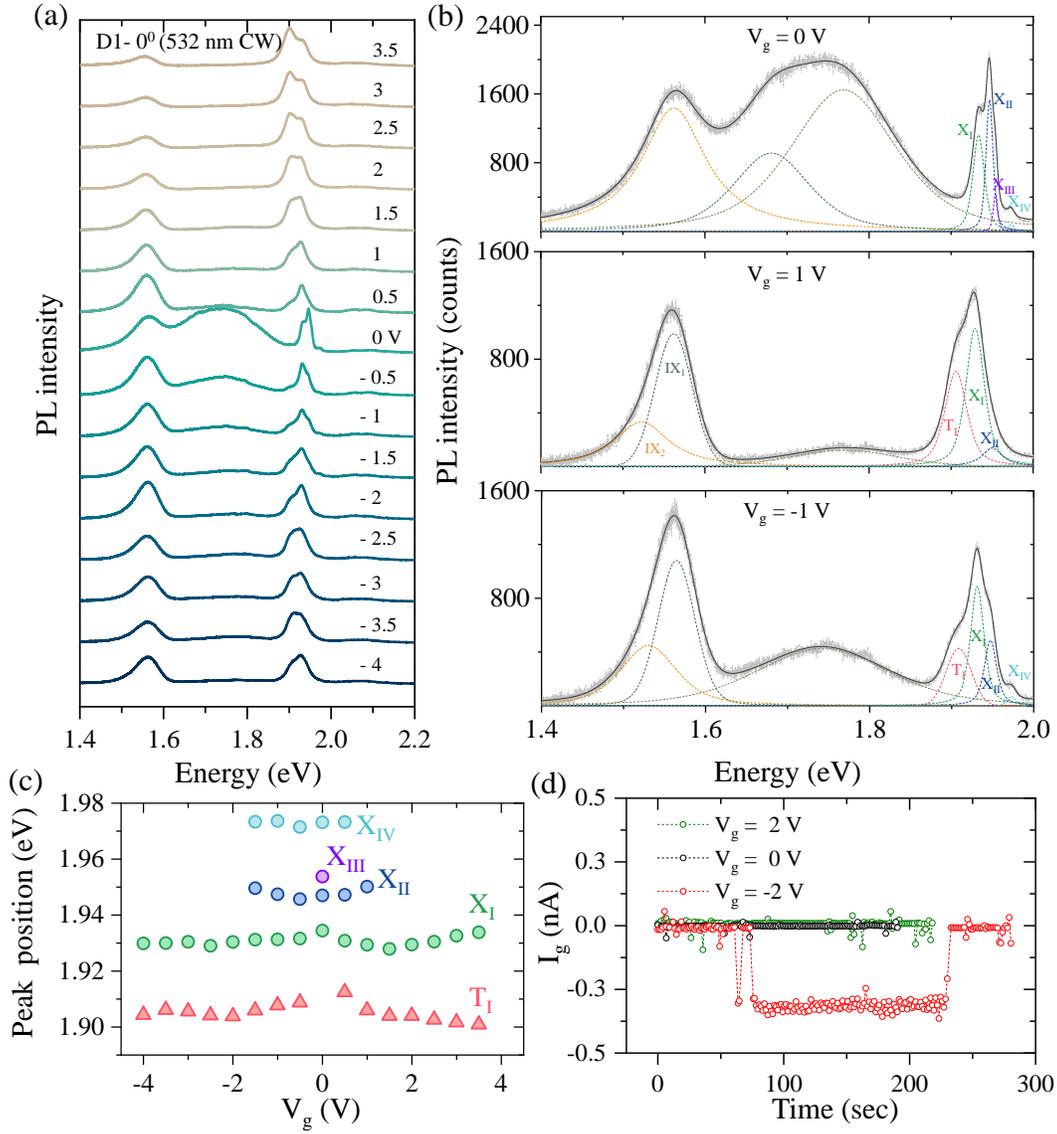

**Figure S4: PL characterization of tBL-MoS$_2$ device-D1.** (a) Individual 532 nm CW laser PL spectra as a function of $V_g$ corresponding to the color plot shown in the main text. (b) Representative fitting of PL spectra at different $V_g$ used for peak analysis. $IX_1$ and $IX_2$ correspond to indirect band emission. (c) Peak position as a function of $V_g$ represented for different peaks obtained from spectral fitting. (d) Time-resolved gate leakage current from D1 at different $V_g$.

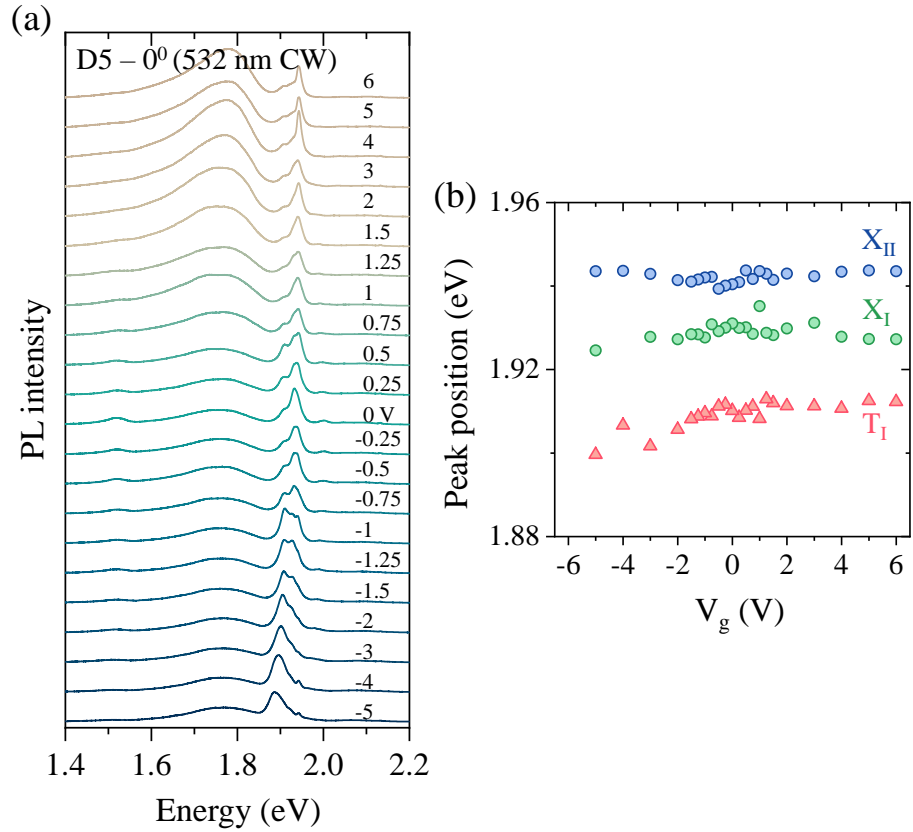

**Figure S5: PL characterization of tBL-MoS$_2$ device-D5.** (a) Gate-dependent PL spectra obtained with 532 nm CW laser from another near-$0^0$ H-type stacked tBL-MoS$_2$ device, D5. (b) Gate-dependent peak positions obtained from the corresponding spectral fitting of the PL data in (a). D5 also exhibits multiple exciton sub-band emission similar to D1 discussed in the main text.

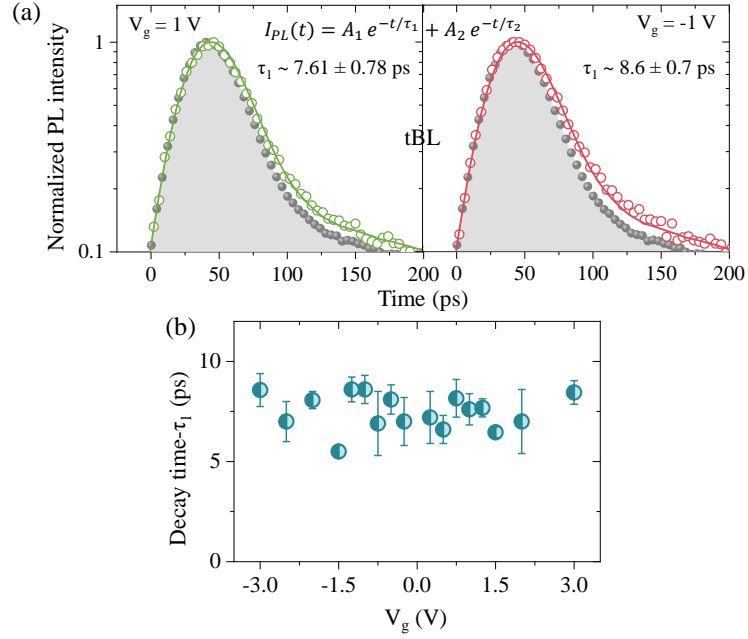

**Figure S6: Exction TRPL and decay analysis of tBL-MoS$_2$ from D1**. (a) Representative fitting (solid lines) from the bi-exponential deconvolution of the exciton TRPL (open circles) at $V_g = +1$ and -1 V highlighting the fastest decay time component ($\tau_1$). IRF is represented in solid grey circles with shaded area. (b) Decay time ($\tau_1$) as a function of $V_g$ extracted from TRPL measurements in the spectral window of moiré excitons.

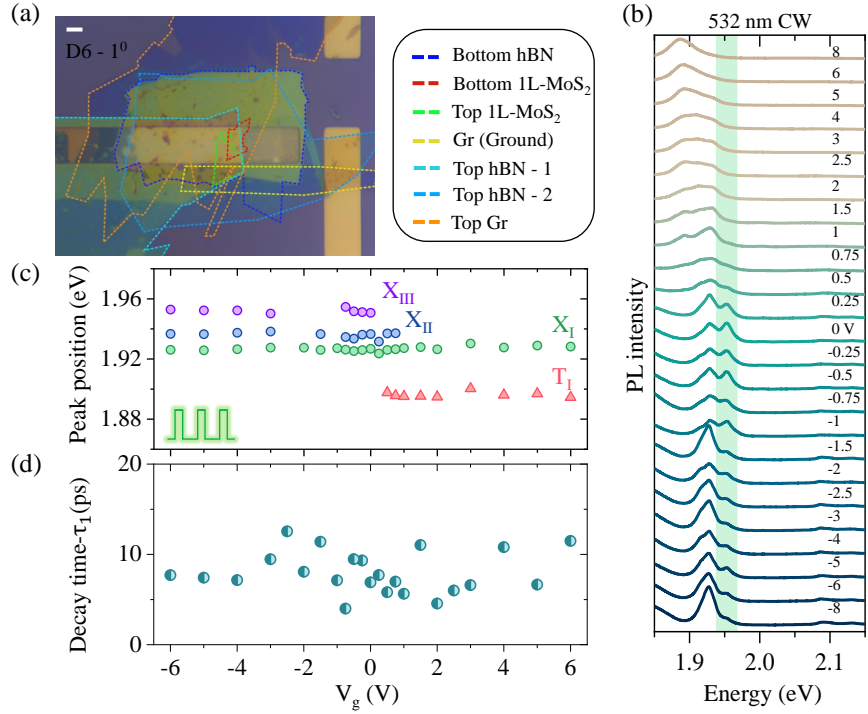

**Figure S7: PL characterization of tBL-MoS$_2$ device-D6.** (a) Optical image of tBL-MoS$_2$ device, D6 with near-$1^0$ twist in H-type stacking. Individual layers are highlighted with colored dashed lines. Scale bar is 5 $\mu$m. (b) Gate-dependent PL spectra with 532 nm CW laser excitation. (c) Peak position variation with gating under 531 nm pulsed laser excitation. (d) Fastest decay component ($\tau_1$) extracted from TRPL data at different gating conditions. The green area in (b) represents the corresponding spectral window of TRPL measurement.

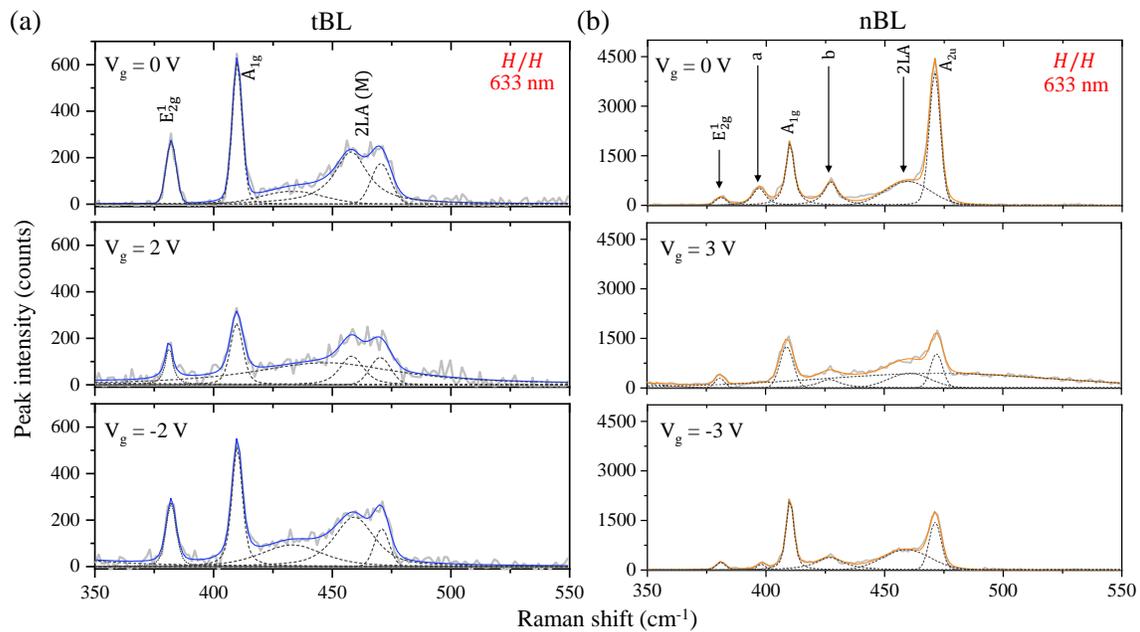

**Figure S8: 633 nm linearly polarized Raman from tBL- and nBL-MoS$_2$.** (a-b) Representative fitting of the Raman peaks from the co-(H/H) polarized spectra at different $V_g$ after appropriate baseline correction for tBL-MoS$_2$ of D1 and nBL-MoS$_2$ of D2 respectively.

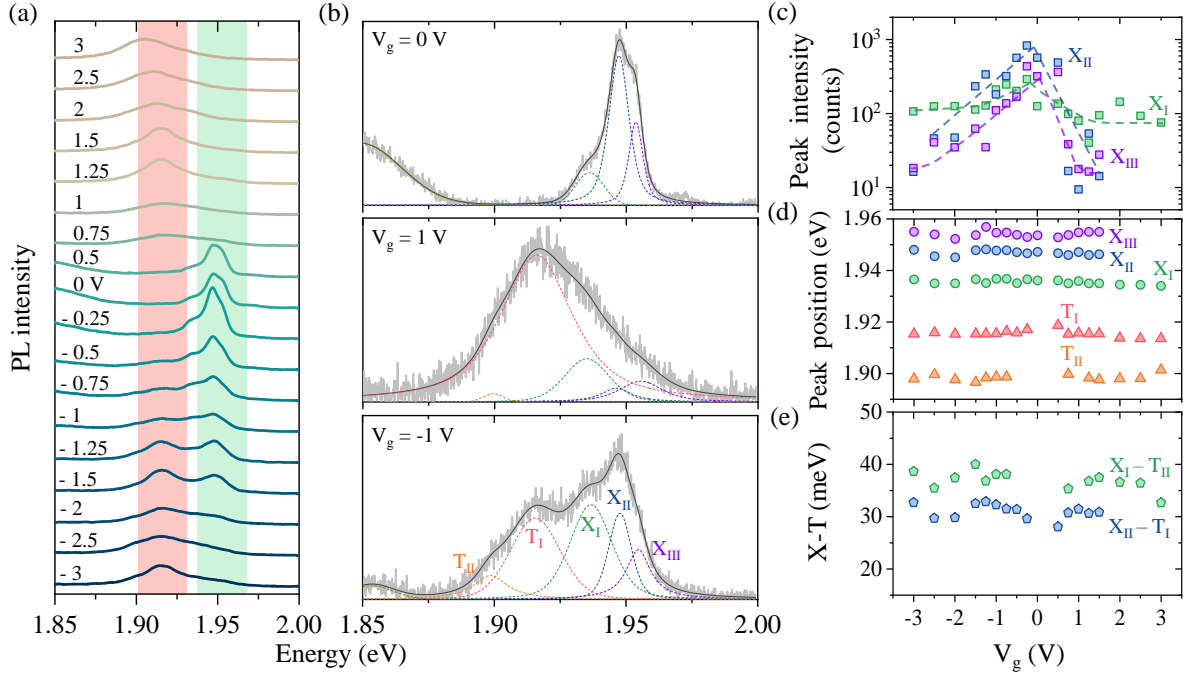

Figure S9: **Pulsed laser excitation of tBL-MoS$_2$ device-D1**. (a) Individual gate-dependent PL spectra corresponding to the color plot shown in the main text. Green and red regions highlight the spectral windows of TRPL measurements for the excitons and trions respectively. (b) Representative fitting of the PL spectra at different $V_g$. A clear evolution of two trion peaks with gating can be seen here. (c-d) Peak intensities and positions extracted from spectral fitting with varying $V_g$. (e) Separation between different exciton and trion peaks at different $V_g$.

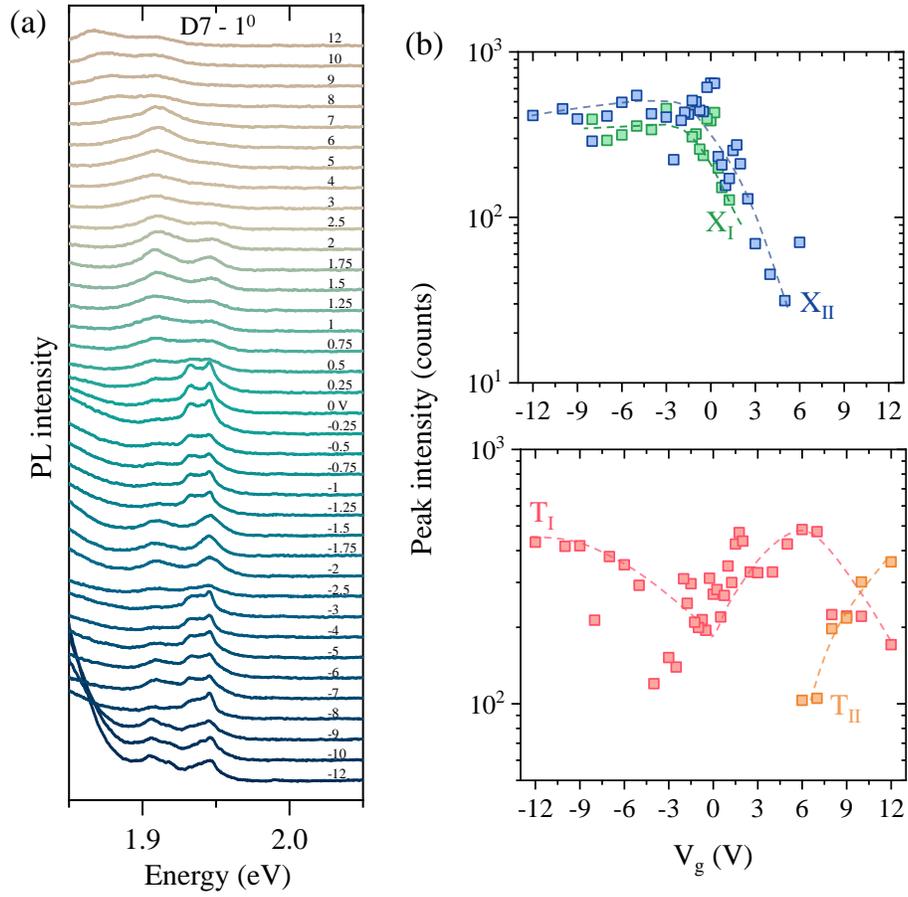

**Figure S10: PL characterization of tBL-MoS$_2$ device-D7.** (a) Gate-dependent PL spectra of another near-1$^0$ tBL-MoS$_2$ in H-type stacking (D7) with 532 nm CW laser excitation. (b) The extracted peak intensities from spectral fitting as a function of $V_g$. Dashed lines are guide to the eye. Evolution of a second trion peak, $T_{II}$ is seen here at very high gating. The intensity of both $T_I$ and $T_{II}$ exhibits a gate dependence similar to D1 discussed in the main text.

# Note-II

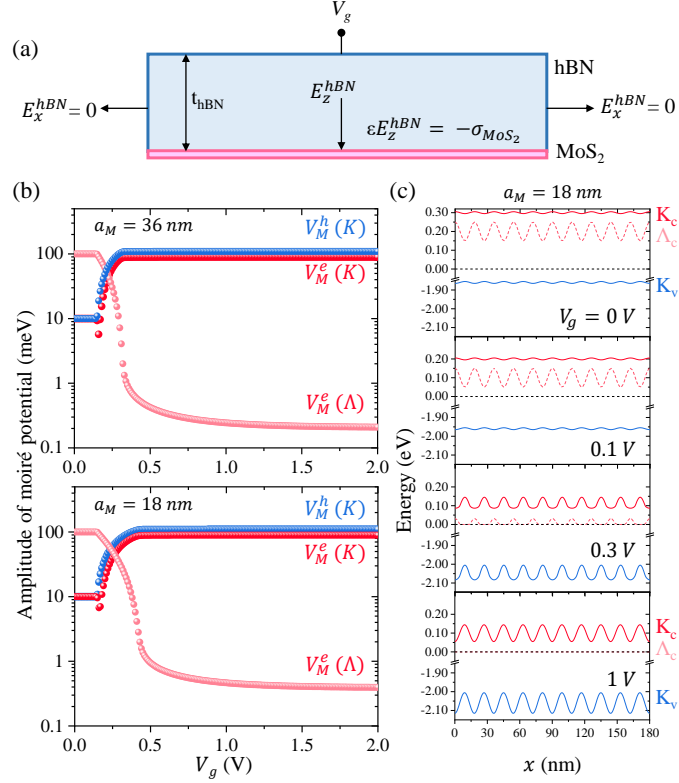

**Figure S11: Electrostatics of tBL-MoS$_2$ from 2D Poisson simulation**. (a) Schematic of the simulation structure used and the corresponding boundary conditions. (b) The amplitude of moiré potential for electrons (red spheres) at $\mathbf{K}$ (dark shaded) and $\mathbf{\Lambda}$ points (light shaded) and holes at $\mathbf{K}$ point (green spheres) as a function of $V_g$ for $a_M$ = 36 nm (top panel) and 18 nm (bottom panel). (c) Spatial profiles of conduction (red) and valence (blue) bands at different $V_g$ for $a_M$ = 18 nm.

To understand the electrostatics of the tBL-MoS$_2$, we numerically solve a 2D Poisson equation ($\nabla^2 \phi = 0$) in the hBN dielectric region with relevant boundary conditions as shown in Fig. S11 (a). The tBL-MoS$_2$ is considered to be a 2D sheet of charge, with a surface charge density $\sigma = q(p - n + N_D)$ where $n$, $p$ and $N_D$ are the electron, hole and impurity 2D surface densities respectively. Carrier densities are calculated using the

standard zeroth order Fermi-Dirac integral as shown below.

$$n(x) = N_{2D} \log\left(1 + \exp\left(\frac{E_F - E_C(x) + q\phi(x)}{k_B T}\right)\right)$$

where $N_{2D}$ is the 2D density of states. Both conduction (CB) and valence band (VB) profiles are assumed to be spatially varying, emulating the spatially varying band gap in a moiré superlattice of tBL-MoS$_2$. CB and VB profiles at the $\Lambda$ and $\Gamma$ points [$E_C^\Lambda(x)$, $E_V^\Gamma(x)$], respectively, corresponding to the indirect band gaps are considered along with CB and VB profiles at $K$ point [$E_C^K(x)$, $E_V^K(x)$] corresponding to the direct band gap. The moiré potential for electrons and holes is assumed to be a simple cosine profile. $E_C^\Lambda(x) = E_{C0}^\Lambda + \Delta E_G^{ind} \cos(k_x x)$ and $E_V^\Gamma(x) = E_{V0}^\Gamma - \Delta E_G^{ind} \cos(k_x x)$ with a spatially varying indirect band gap of $E_G^{ind}(x) = (E_{C0}^\Lambda - E_{V0}^\Gamma) + 2\Delta E_G^{ind} \cos(k_x x)$ while $E_C^K(x) = E_{C0}^K + \Delta E_G^d \cos(k_x x)$ and $E_V^K(x) = E_{V0}^K - \Delta E_G^d \cos(k_x x)$ with a spatially varying direct band gap, $E_G^d(x) = (E_{C0}^K - E_{V0}^K) + 2\Delta E_G^d \cos(k_x x)$. $k_x = \frac{2\pi}{a_M}$, where $a_M$ is the period of Moiré potential. We performed simulations for both $a_M$ = 18 nm ($\theta = 1^0$) and 36 nm ($\theta = 0.5^0$). $\Delta E_G^{ind}$ and $\Delta E_G^d$ are assumed to be 50 and 5 meV, respectively, for both $\theta = 0.5^0$ and $1^0$. $\Delta E_G^{ind}$ is much stronger than $\Delta E_G^d$ due to stronger hybridization effects at the indirect band gap.[1] Temperature (T) used is 4 K. Valley degeneracy of 1, 2 and 6 are considered at the $\Gamma$, $K$ and $\Lambda$ points, respectively. The electronic direct and indirect band gaps (in the absence of moiré potential) are assumed to be 2.16 and 1.55 eV respectively. $E_{C0}^K - E_{C0}^\Lambda$ is taken as 0.1 eV. The thickness ($t_{hBN}$) and dielectric constant of hBN considered are 15 nm and 5 respectively. Effective mass for electrons and holes is assumed to be 0.45 $m_0$. The tBL-MoS$_2$ is considered to be $n$-doped initially at $V_g = 0$ V with the Fermi-level ($E_F$) 0.3 eV below the $E_{C0}^\Lambda$.

Fig. S11(b) shows the variation in the amplitude of moiré potential with positive gating

for both electrons and holes at different points of momentum for $a_M = 36$ and 18 nm. The moiré potential for electrons at $\boldsymbol{\Lambda}$ point gradually weakens after certain $V_g$ where the $E_F$ moves very close to $E_C^\Lambda$ [Fig. S11(c)] while the moiré potential for electrons and holes at the $\boldsymbol{K}$ point increases. However, the amplitude of the moiré potential of electronic direct or indirect band gap is same for all $V_g$.

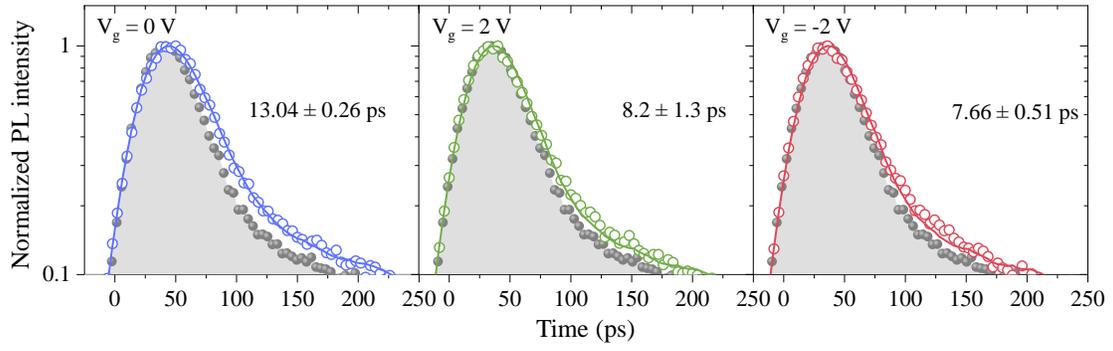

**Figure S12: Trion TRPL and decay analysis of tBL-MoS$_2$ from D1**. Representative fitting (solid lines) from the bi-exponential deconvolution of the trion TRPL (open circles) at different $V_g$ highlighting the fastest decay time component ($\tau_1$). IRF is represented in solid grey circles with shaded area.

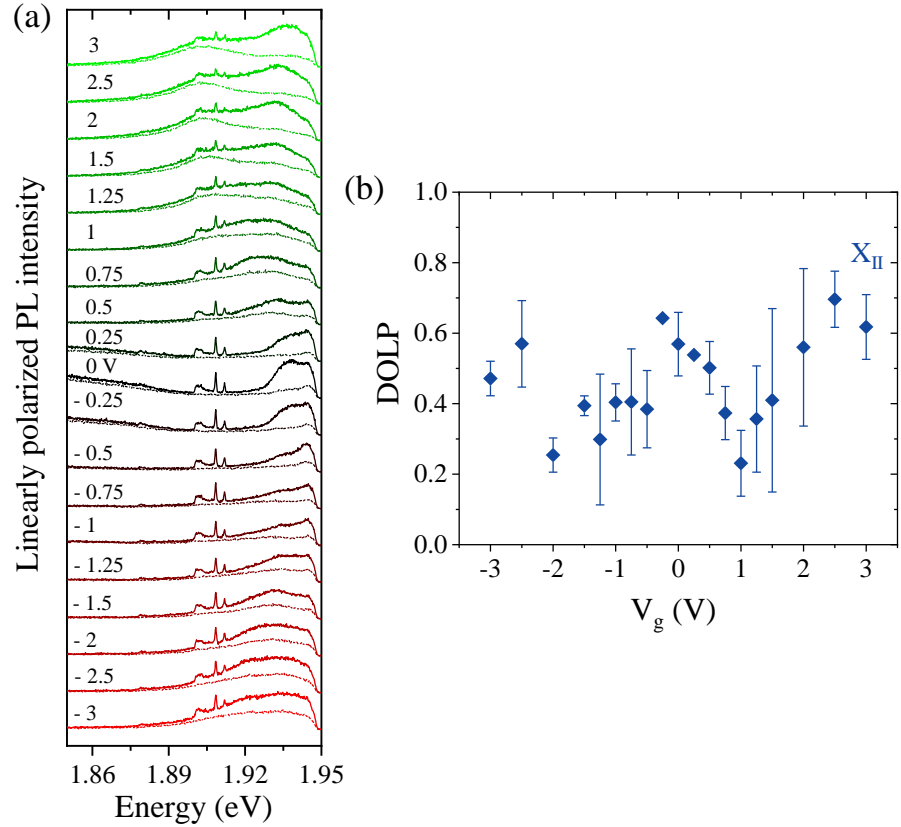

**Figure S13: 633 nm linearly polarized PL from tBL-MoS$_2$ of D1**. (a) Individual co- (solid) and cross-polarized (dotted) line spectra with near-resonant 633 nm linearly polarized excitation. (b) Experimental trend of degree of linear polarization (DOLP) of the delocalized exciton ($X_{II}$) with varying $V_g$.


\end{document}